\documentclass[]{elsarticle}

\makeatletter
\def\ps@pprintTitle{%
 \let\@oddhead\@empty
 \let\@evenhead\@empty
 \def\@oddfoot{LLNL-JRNL-829386-DRAFT}%
 \let\@evenfoot\@oddfoot}
\makeatother

\usepackage{lineno,hyperref}
\modulolinenumbers[5]

\newcommand{\aoc}{\textsf{A1C}}
\newcommand{\crp}{\textsf{CRP}}
\newcommand{\chol}{\textsf{Cholesterol}}
\newcommand{\diab}{\textsf{250.XX}}
\newcommand{\phqt}{\textsf{PHQ-2}}
\newcommand{\phqn}{\textsf{PHQ-9}}

\usepackage{amsmath}
\usepackage{algorithm}
\usepackage{algpseudocode}
\usepackage{caption}
\usepackage{subcaption}

\journal{Journal of Biomedical Informatics}

\begin{document}

\begin{frontmatter}

\title{Continuous-Time Probabilistic Models for Longitudinal Electronic Health Records}


\author[llnl]{Alan D. Kaplan\corref{mycorrespondingauthor}}
\cortext[mycorrespondingauthor]{Corresponding author}
\ead{kaplan7@llnl.gov}

\author[llnl]{Uttara Tipnis}
\author[durham,vama,duke]{Jean C. Beckham}
\author[durham,vama,duke,vahsnc]{Nathan A. Kimbrel}
\author[visn,penn]{David W. Oslin}
\author[]{the MVP Suicide Exemplar Workgroup}
\author[lanl]{Benjamin H. McMahon}

\address[llnl]{Computational Engineering Division, Lawrence Livermore National Laboratory, 7000 East Ave., Livermore, CA 94550}
\address[durham]{Durham Veterans Affairs (VA) Health Care System, Durham, NC, USA}
\address[vama]{VA Mid-Atlantic Mental Illness Research, Education and Clinical Center, Durham, NC, USA}
\address[duke]{Department of Psychiatry and Behavioral Sciences, Duke University School of Medicine, Durham, NC, USA}
\address[vahsnc]{VA Health Services Research and Development Center of Innovation to Accelerate Discovery and Practice Transformation, Durham, NC, USA}
\address[visn]{VISN 4 Mental Illness Research, Education, and Clinical Center, Center of Excellence, Corporal Michael J. Crescenz VA Medical Center, Philadelphia, PA, USA}
\address[penn]{Department of Psychiatry, Perelman School of Medicine, University of Pennsylvania, PA, USA}
\address[lanl]{Theoretical Biology and Biophysics, Los Alamos National Laboratory, Los Alamos, NM, USA}

\begin{abstract}
Analysis of longitudinal Electronic Health Record (EHR) data is an important goal for precision medicine.
Difficulty in applying Machine Learning (ML) methods, either predictive or unsupervised, stems in part from the heterogeneity and irregular sampling of EHR data.
We present an unsupervised probabilistic model that captures nonlinear relationships between variables over continuous-time.
This method works with arbitrary sampling patterns and captures the joint probability distribution between variable measurements and the time intervals between them.
Inference algorithms are derived that can be used to evaluate the likelihood of future using under a trained model.
As an example, we consider data from the United States Veterans Health Administration (VHA) in the areas of diabetes and depression.
Likelihood ratio maps are produced showing the likelihood of risk for moderate-severe vs minimal depression as measured by the Patient Health Questionnaire-9 (PHQ-9).
\end{abstract}


\end{frontmatter}


\section{Introduction}
Improved individualized patient care is a central goal of precision medicine \cite{Ginsburg2018-zf, Kosorok2019-bz}.
Historical records containing clinical measurements, laboratory results, diagnoses, and outcomes offer an opportunity to learn about the individual characteristics that lead to increased risk of disease or disorder.
Using such records may help to inform the development of more precise and individualized treatment.
Large quantities of Electronic Health Records (EHRs) have been collected in part to help with the development of data-driven precision medicine methods and systems \cite{Kim2019-bo}.

One limitation that has been studied in this context is the development of data-driven prediction of outcomes, such as mortality.
This can be done in either acute (e.g. emergency department, inpatient, ICU), or outpatient settings.
Many machine learning (ML) techniques have been applied toward such prediction problems, such as decision trees and random forests \cite{Weiss2012-qp, Ellis2019-hs, Rahimian2018-ih, Xie2020-jl, Zhou2016-po, Ye2020-zx}, neural networks \cite{Kam2017-tz, Rasmy2018-gi}, and regression techniques \cite{Su2020-sq, Levine2018-ih, Wu2010-hr}.
Rather than computing an outcome prediction, probabilistic unsupervised methods are geared toward a related, but different goal \cite{Ghahramani2015-xn, Murphy2012-kt}.
The aim of these methods is to construct a probability distribution model of the data that quantifies the likelihood of a collection of variables.
This model can then be used to perform inference and compute probabilities for uncertain events.

However, EHR data have irregularities that make applying ML methods difficult \cite{Shickel2018-qy}.
EHR data contain a large number of variables, each of which can be sampled irregularly.
In addition, the data types are heterogeneous, containing discrete, ordinal, and continuous value variables.
Many existing methods address these challenges by transforming the data into a form that is amenable to application of the ML methods, such as time-windowing and quantization.
However, these processes result in a typically unknown loss of information.
Using a constant time window across variables may also be problematic since different time scales may apply to different variables.

In this work, we address these challenges by constructing continuous-time unsupervised probabilistic models.
The model is a pairwise joint probability distribution between two measured variables and the time interval between measurements.
It can be trained on longitudinal data containing arbitrary sample patterns in a computationally scalable manner, and does not use any lossy transformations of the data or time-windowing.
In this way, all available timepoints can be used in estimating the model parameters.
Once trained, the model can be used to compute probabilities of events of interest, such as the odds ratio of an outcome as a function of time.
In addition, multiple models can be composed to form a model of an entire EHR collection.

Predictive methods are geared towards accurate prediction of a target variables.
Several approaches have been developed for EHR data that operate on longitudinal data streams.
Neural network models for sequence data, such as Long-Short-Term Memory (LSTM) and Gated Recurrent Units (GRU) have been used to incorporate the temporal effects of EHR data in predictive methods.
These can be applied in discretized steps \cite{Jin2018-rh}, or directly operate on time intervals \cite{Choi2016-pr}.
These recurrent neural networks track the state dynamics and generate representations of the data at each step that can be used to predict outcomes.
Rather than a predictive approach, we focus on probabilistic representations of data that can be used to compute statistical likelihoods of various outcomes with a single model.

Probabilistic models trainable on sequential EHR data have also been developed.
Markov-based modeling has been used to characterize state transitions over time \cite{Huang2018-oy}.
Continuous-time versions of these probabilistic methods include the continuous-time Markov chain and continuous-time hidden Markov model.
These methods and continuous-time Bayesian networks model the time intervals between measurements in a Markov state sequence \cite{Liu2015-ad, Stella2012-vr}.
The present model is designed to capture the joint distribution between time intervals and measurements, rather than the transitions between timepoints.

A related set of techniques based on probabilistic topic models can be used to uncover underlying patterns in an unsupervised model.
This joint probability modeling in the form of graphical models can be used to model heterogeneous data types \cite{Pivovarov2015-zu, Mayhew2018-pt}.
These approaches incorporate the use of a latent variable to control conditional dependence between data elements, and can be used to derive data-driven phenotypes.
Versions of Latent Dirichlet Allocation (LDA) have been applied to EHR data \cite{Huang2015-dq}.
Similar to these methods, our approach is a latent variable model, however it models continuous time intervals explicitly.

Survival modeling is another related set of approaches for predicting the time to an event.
The Cox proportional hazards model is a regression-based approach that can be used to estimate risk over time given a fixed set of covariates (see e.g. \cite{Ohno-Machado2001-ax}).
Versions incorporating time-varying regression coefficients allow for covariates to have time-dependent effects.
Adaptations for longitudinal EHR data with irregularly sampled response variables can be formulated as a marked point process \cite{Martinussen2006-iw}.
Covariates can reflect internal dynamics such as the time since the last measurement.
These methods generate partial likelihood functions, unlike the full joint probability distributions that the method in this paper describes.

In this work we used data from the Department of Veteran Affairs (VA) electronic medical record.
The VA is the largest health care system in the US comprising of more than 150 medical centers and more than 1000 community based outpatient clinical sites\footnote{https://www.va.gov/health/}.
The VA EHR has been in existing from the late 1990’s.
For this project we used data from 2000 to 2019.


\section{Methods} \label{sec:methods}
\subsection{Continuous-time Model} \label{sec:methods_model}
\begin{figure}[]
    \centering
    \includegraphics[width=0.5\textwidth]{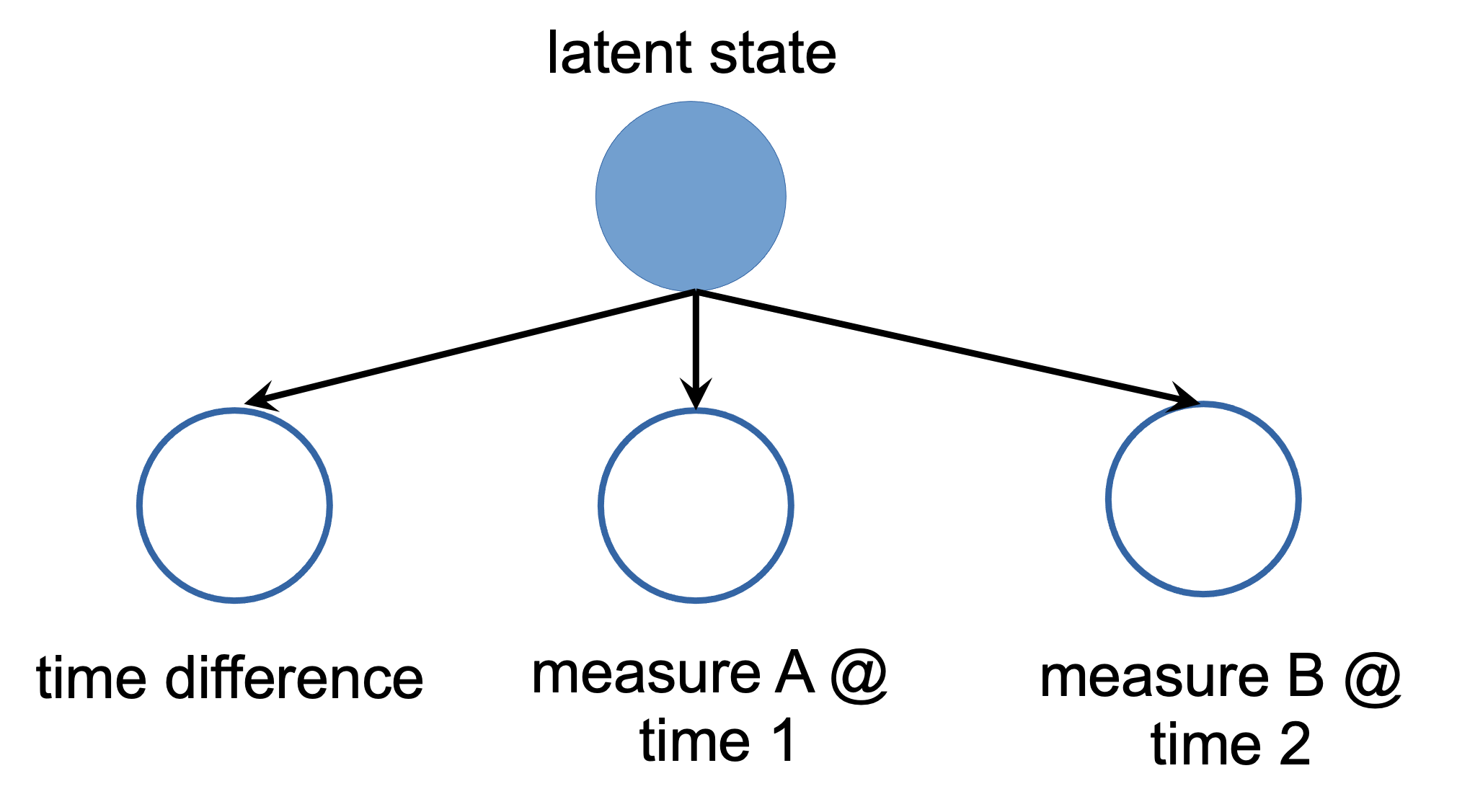}
    \caption{Graphical representation of the model, which is a mixture across two measurements and the time interval between them. Arrows indicate conditional dependence structure in the joint probability distribution.}
    \label{fig:model}
\end{figure}

\paragraph{Model definition}
Our method is based on estimating the joint probability distribution between two variables and the time that has elapsed between their measurement.
The model can be viewed as a latent variable graphical model (Figure \ref{fig:model}).
The density function is a mixture of distributions,
    \begin{equation} \label{eq:modeldef}
        f\left(\delta, x, y; N_Z\right) = \sum_{z=1}^{N_Z}
        \alpha_z
        f_{exp}\left(\delta; \lambda_z\right)
        f_G\left(x; \mu_z, \sigma_z^2\right)
        f_G\left(y; \nu_z, \xi_z^2\right),
    \end{equation}
where $\delta$ is the time difference, and $x$ and $y$ are the variables.
Within the mixture, the time difference is characterized by exponential distributions $f_{exp}(\delta;\lambda)$, and the two variables are characterized by Gaussian distributions $f_G(x;\mu, \sigma^2)$.
Mixing parameters are $\alpha_z \ge 0$ with $\sum_{z=1}^{N_Z} \alpha_z = 1$.
The number of components is $N_Z$ and the total parameter count for the model is $p=6N_Z - 1$.
See \ref{app:not} for a description of the notation used in this work.

\paragraph{Estimation}
Estimation of the parameters is performed using Expectation Maximization (EM) (see e.g. \cite{Moon1996-kb}).
Given $N$ samples $(\delta^{(1)}, x^{(1)}, y^{(1)}), \ldots (\delta^{(N)}, x^{(N)}, y^{(N)})$, we use the EM procedure to estimate the parameters: $\boldsymbol{\alpha}, \boldsymbol{\lambda}, \boldsymbol{\mu}, \boldsymbol{\sigma}^2, \boldsymbol{\nu}, \boldsymbol{\xi}^2$ that maximize $f(\boldsymbol{\delta}, \boldsymbol{x}, \boldsymbol{y};N_Z) = \prod_{i=1}^N f(\delta^{(i)}, x^{(i)}, y^{(i)};N_Z)$.
The EM algorithm for this model is shown in Algorithm \ref{alg:est}.
See \ref{app:est} for details on how this is derived.

Convergence of the estimation algorithm can be performed by either examining the parameters or the likelihood function.
In our experiments we keep track of the log-likelihood per sample for every iteration.
We claim convergence if the fractional increase in the fit is less than a fixed value (0.01) for all of the last 10 iterations.

\begin{algorithm}
    \caption{Parameter Estimation}\label{alg:est}
    \begin{algorithmic}
    \State \bf{Initialize} $\boldsymbol{\alpha} \gets \boldsymbol{\alpha}^{(0)}$, 
    $\boldsymbol{\lambda} \gets \boldsymbol{\lambda}^{(0)}$, 
    $\boldsymbol{\mu} \gets \boldsymbol{\mu}^{(0)}$, 
    $\boldsymbol{\sigma}^2 \gets \boldsymbol{\sigma}^{2(0)}$, 
    $\boldsymbol{\nu} \gets \boldsymbol{\nu}^{(0)}$, 
    $\boldsymbol{\xi}^2 \gets \boldsymbol{\xi}^{2(0)}$
    \State $j \gets 0$
    \While{not converged}
        \State $\gamma_i(z) \gets \frac{\alpha_z^{(j)}
        f_{exp}\left(\delta_i; \lambda_z^{(j)}\right)
        f_G\left(x_i; \mu_z^{(j)}, \sigma_z^{2(j)}\right)
        f_G\left(y_i; \nu_z^{(j)}, \xi_z^{2(j)}\right)}{f\left(\delta_i, x_i, y_i\right)}$
        \State $\gamma(z) = \sum_{i=1}^n \gamma_i(z)$
        \State $\alpha_z^{(j+1)} \gets \frac{\gamma(z)}{n}$
        \State $\lambda_z^{(j+1)} \gets \frac{\gamma(z)}{\sum_{i=1}^n \gamma_i(z)\delta_i}$
        \State $\mu_z^{(j+1)} \gets \frac{\sum_{i=1}^n \gamma_i(z)x}{\gamma(z)}$
        \State $\sigma_z^{2(j+1)}\gets \frac{\sum_{i=1}^n \gamma_i(z)(x-\mu_z)^2}{\gamma(z)}$
        \State $\nu_z^{(j+1)} \gets \frac{\sum_{i=1}^n \gamma_i(z)x}{\gamma(z)}$
        \State $\xi_z^{2(j+1)}\gets \frac{\sum_{i=1}^n \gamma_i(z)(x-\nu_z)^2}{\gamma(z)}$
        \State $j \gets j + 1$
    \EndWhile
    \end{algorithmic}
\end{algorithm}

\paragraph{Model selection}
The number of components, $N_Z$, is a parameter that controls the model complexity.
A greater $N_Z$ increases the complexity and expressiveness of the model, but reduces its generalizability.
Selecting an appropriate value for $N_Z$ balances model complexity while reducing overfitting of the model to training data.
We use the Bayesian Information Criterion (BIC) as a target to optimize $N_Z$ (see e.g. \cite{Hastie2009-ba}).
The BIC for our model is $BIC(N_Z)=(6N_Z - 1)\log N - 2\log f(\boldsymbol{\delta}, \boldsymbol{x}, \boldsymbol{y}; Z)$.
The goal is to find $N_Z$ that minimizes the BIC score.
Although it is possible to perform a linear search over $N_Z$, in some cases this can be an overly expensive approach.
In our experiments we utilize Bayesian Optimization (BO) methods to optimize the BIC \cite{Snoek2012-qk}.
This established approach estimates a Gaussian Process to characterize the uncertainty of the $BIC(N_Z)$ function and selects points to sample that satisfy a defined criterion.
For model selection, the BO operates on an outer loops, with EM estimation occurring in the inner loop.
Every outer iteration estimates a new model with a value of $N_Z$ chosen by the BO.
The model with the lowest BIC is chosen as the final model.
See \ref{app:bo} for details on parameters used for this approach.

\paragraph{Inference algorithms}
Once a model is trained, there are a number of inference procedures that can be applied.
The distribution of one variable given another over time can be computed by calculating
    \begin{equation} \label{eq:inf}
        f\left(y|x,\delta\right) = \sum_{z=1}^{N_Z} \Pr\left(Z=z|x,\delta\right) f_G\left(y;\nu_z, \xi_z^2\right),
    \end{equation}
which is a Gaussian Mixture Model with mixing coefficients determined by the inputs $x$ and $\delta$.
The distribution of the latent variable given $x$ and $\delta$ can be found using
    $$
        \Pr\left(Z=z|x,\delta\right) = 
        \frac{\alpha_z
        f_{exp}\left(\delta; \lambda_z\right)
        f_G\left(x; \mu_z, \sigma_z^2\right)}
        {f\left(x, \delta\right)}.
    $$
The normalizing constant in the denominator does not depend on $z$ and ensures that $\sum_z \Pr\left(Z=z|x,\delta\right) = 1$.

More than one model can be composed to perform inference on a common variable.
In this setup, we would like to compute the likelihood of a target variable $Y$ given
input variables $X_1, \ldots, X_M$ and time differences between these variables and $Y$, $\Delta_1, \ldots, \Delta_M$.
This can be performed using 
a collection of models $f_1, \ldots, f_M$, where $f_i(\delta, x, y)$ is the joint distribution between $(\Delta_i, X_i, Y)$.
Composing these models together yields
    \begin{equation} \label{eq:comp}
        f(y|x_1, \ldots x_M, \delta_1, \ldots, \delta_M) = \prod_{i=1}^M f\left(y|x_i, \delta_i\right).
    \end{equation}
This framework can be used to infer future values of $Y$ given past values of the input variables.
If the current time is $t=0$ and the input variables were collected at time $t_1,\ldots,t_m$, then using the trained models the distribution of $Y$ at time $t$ is $f(y|x_1, \ldots x_m, t - t_1, \ldots, t - t_m).$

\subsection{Baseline Models}
We formulate two baseline models.
These are used to evaluate and compare the fit of data to our model.
The first model is a multivariate Gaussian:
    \begin{equation} \label{eq:baseline1}
        f\left(\delta, x, y\right) = 
        \frac{1}{\sqrt{2\pi|\boldsymbol{\Sigma|}}}
        e^{-\frac{1}{2}
        \left([\delta, x, y] - [\mu_\delta, \mu_x, \mu_y]\right)^T
        \boldsymbol{\Sigma}^{-1}
        \left([\delta, x, y] - [\mu_\delta, \mu_x, \mu_y]\right)
        },
    \end{equation}
where $\mu_\delta$, $\mu_x$, and $\mu_y$ are the mean parameters, and $\boldsymbol{\Sigma}$ is the covariance matrix.

The second baseline model is a conditional bivariate Gaussian model, where a bivariate Gaussian is conditioned on the value of an ordinal variable $y$.
The probability density function for this model is,
    \begin{equation} \label{eq:baseline2}
        f\left(\delta, x, y\right) = 
        p_y
        \frac{1}{\sqrt{2\pi|\boldsymbol{\Sigma|}_y}}
        e^{-\frac{1}{2}
        \left([\delta, x] - [\mu_{\delta|y}, \mu_{x|y}]\right)^T
        \boldsymbol{\Sigma}_y^{-1}
        \left([\delta, x] - [\mu_{\delta|y}, \mu_{x|y}]\right)
        },
    \end{equation}
where $\mu_{delta|y}$, $\mu_{x|y}$ are the mean parameters dependent on $y$, and $\boldsymbol{\Sigma}_y$ is the covariance matrix dependent on $y$.
This model allows for increased expressiveness compared to (\ref{eq:baseline1}), however it has the requirement that $y$ is ordinal.

\subsection{Variables and Data Extraction} \label{sec:methods_vars}
In the following case studies, we use the following variables:
\begin{itemize}
    \item \aoc{} is the hemoglobin A1C level as recorded in a standard laboratory test. The normal range is less than 5.7\%, 5.7\%--6.4\% indicates prediabetes, and greater than 6.5\% indicates diabetes\footnote{https://www.cdc.gov/diabetes/managing/managing-blood-sugar/a1c.html}.
    \item \chol{} is the total cholesterol as recorded in a standard laboratory test. The normal range is less than 200 mg/dL, 200 mg/dL - 239 mg/dL is borderline high, and greater than 240 mg/dL is considered high.
    \item \diab{} is any ICD9 code that starts with 250, indicating a diabetes diagnosis,
    \item \crp{} is the level of C-Reactive Protein as recorded in a standard laboratory test, which is a measure of inflammation,
    \item \phqt{} is the 2-item Patient Health Questionnaire used as a screening tools for depressed mood,
    \item \phqn{} is the 9-item Patient Health Questionnaire used to assess depression severity on a 0-27 point scale, with high values indicating greater severity.
\end{itemize}
Using the VHA patient record data, we organize the subjects into 100 cohorts based on age.
Each cohort contains approximately 230,000 patients.
For each case study, we randomly sample 100 patients from each cohort that have at least one recorded value from each variable.
We then remove instances that have incomplete recordings, such as laboratory values missing.

\section{Results} \label{sec:res}
In this section, we describe results on synthetic examples, diabetes, and depression.
Synthetic examples (Section \ref{sec:synth}) are designed to demonstrate the ability to model nonlinear and time-dependent behavior.
Results on diabetes data (Section \ref{sec:res_diab}) are based on well-known relationships between hemoglobin A1C and diabetes, and is used to verify that the approach can capture previously understood relationships.
Depression (Section \ref{sec:res_dep}) focuses on much weaker and less understood relationships between laboratory results and survey-based depression tools.
Table \ref{tab:nums} shows the number of subjects, samples, and components that resulted from model selection for all models.
\phqn{} models were trained by performing 20 runs of the model selection procedure with different random initializations for each one and choosing the model with the best BIC value.
Over the 20 runs, we show the mean and 2 standard deviations of the model fit in Table \ref{tab:nums}.

\subsection{Synthetic Examples} \label{sec:synth}
The models for these synthetic examples were designed to express time-dependent interaction between two variables $X$ and $Y$.
After defining the models, we then sample from them and re-estimate the parameters.
Then the expected value of $Y$ given $X$ is computed over time to illustrate the association between variables.
This is compared to the same computation for the estimated model to visualize estimation accuracy.

\paragraph{Diminishing temporal effect} \label{sec:synth1}
In the first example, we construct a model to show a positive association between two variables $X$ and $Y$ that decreases in strength over time.
This is designed to represent commonly occurring behavior.
Table \ref{tab:syn1} shows the parameters chosen to illustrate this relationship between $X$ and $Y$.
The 6 components in this model all have equal mixing coefficients.
The mean of the exponential distribution ($1/\lambda_z$) for each component controls its temporal reach.
Relative to the other components, components 1 and 2 have greater influence in earlier times, while components 5 and 6 have weaker influence later in time.
All variance parameters are set to 1.

\begin{table}[]
  \begin{center}
    \caption{Synthetic Model Showing Diminishing Temporal Effect}
    \label{tab:syn1}
    \begin{tabular}{c||c|c|c|c}
      z & $\alpha_z$ & $\lambda_z$ & $\mu_z$ & $\nu_z$ \\
      \hline
      1 & 0.17 & 1.00 & 0.0 & 0.0 \\
      2 & 0.17 & 1.00 & 1.0 & 1.0 \\
      3 & 0.17 & 0.50 & 0.0 & 0.0 \\
      4 & 0.17 & 0.50 & 1.0 & 0.5 \\
      5 & 0.17 & 0.33 & 0.0 & 0.0 \\
      6 & 0.17 & 0.33 & 1.0 & 0.2 \\
    \end{tabular}
  \end{center}
\end{table}

\begin{figure}[]
    \centering
    \begin{subfigure}[]{0.49\textwidth}
        \centering
        \includegraphics[width=\textwidth]{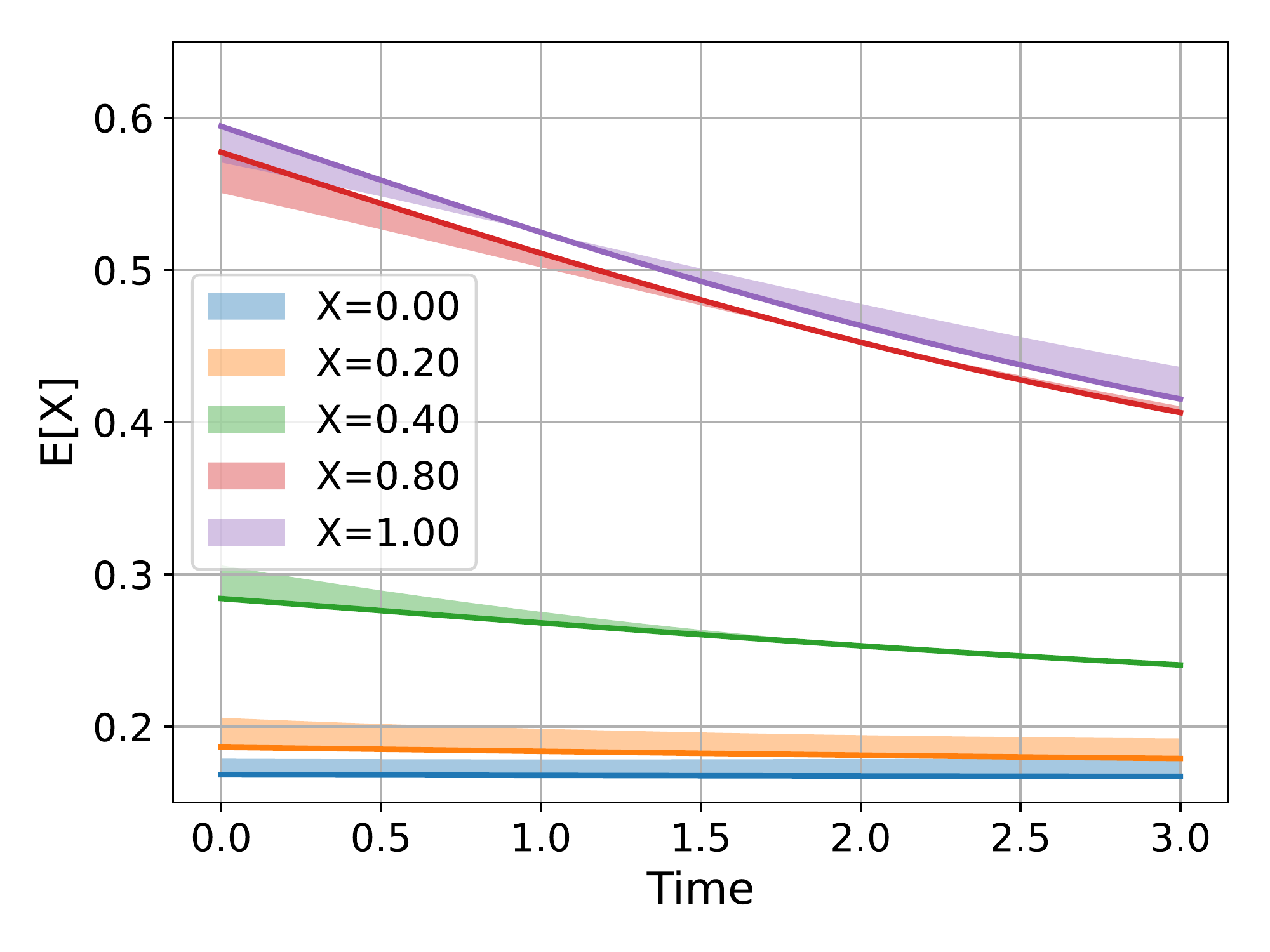}
        \caption{}
        \label{fig:synth1_time}
    \end{subfigure}
    \hfill
    \begin{subfigure}[]{0.49\textwidth}
        \centering
        \includegraphics[width=\textwidth]{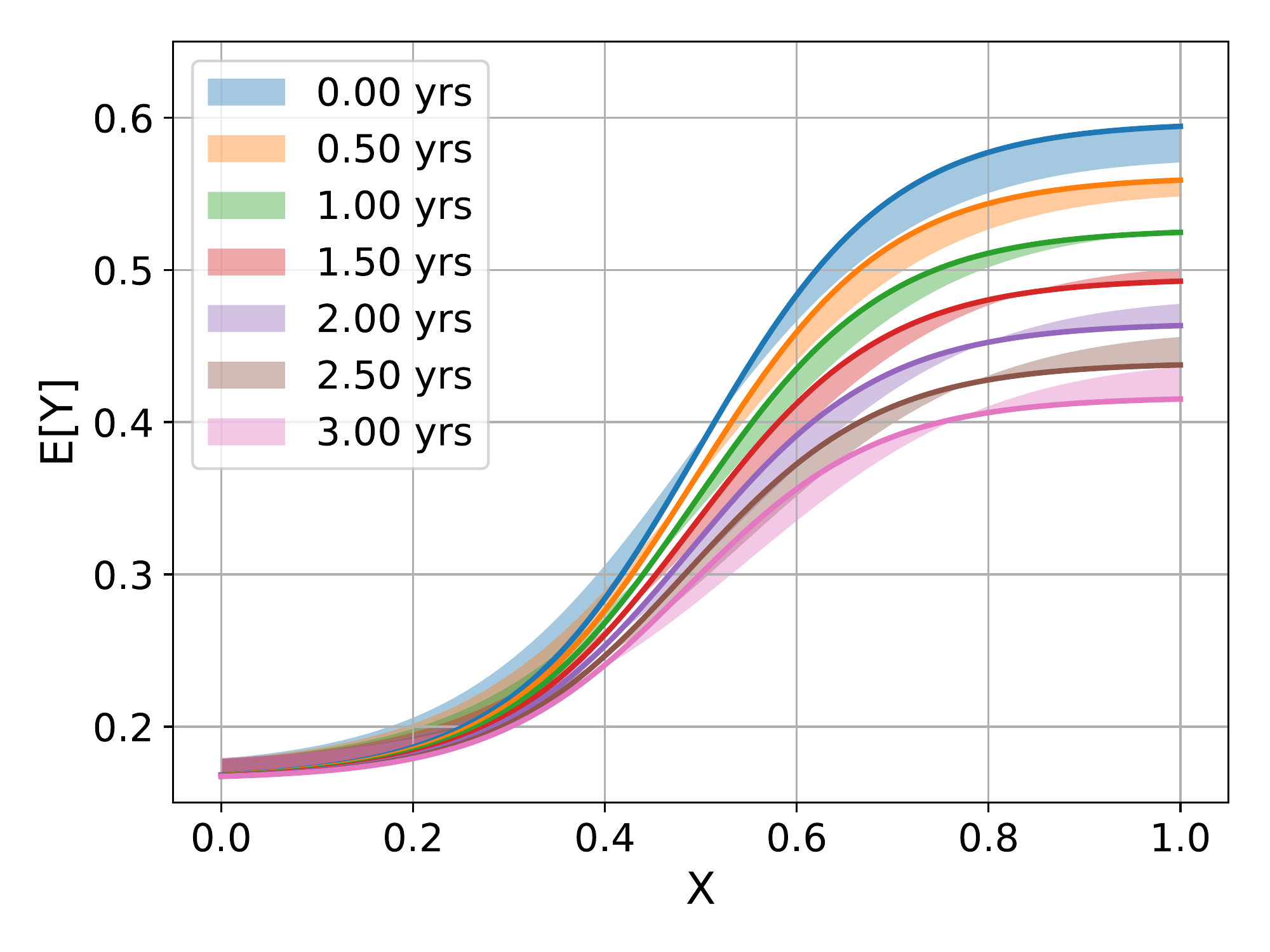}
        \caption{}
        \label{fig:synth1_x}
    \end{subfigure}
    \caption{Inference results for synthetic example with diminishing positive association (Section \ref{sec:synth1}). Figure \ref{fig:synth1_time} shows the expected value of $Y$ over time for selected values of $X$. Figure \ref{fig:synth1_x} shows the same inference, but as a function of continuously valued $X$ with selected time points. For both plots, solid lines are from the original model, and shaded regions are deviations to the re-estimated model.}
    \label{fig:synth1}
\end{figure}

The expected value of $Y$ given $X$ over time, $E[Y|X=x,\Delta=\delta]$ is shown in Figure \ref{fig:synth1_time}.
This is computed by taking the expected value of the distribution in (\ref{eq:inf}), $\sum_y yf\left(y|x,\delta\right)$.
The solid lines are the expected values using the original model, and the shaded area is the region between the original model and a re-estimated model, trained on 10,000 samples drawn from the model.
The same computation, but plotted as the expected value as a function of $x$ is shown in Figure \ref{fig:synth1_x}.
As time increases, the effect that $X$ has on $Y$ decreases.

\paragraph{Changing directionality over time} \label{sec:synth2}
In the second example, $X$ has a positive association with $Y$ initially, but over time inverts to a negative effect.
Table \ref{tab:syn2} shows the parameters used in this 8 component model.
As in the previous example, all components have equal mixing coefficients.
The means of the time intervals in this model ($1/\lambda_z$) are 1, 1.5, 2, and 3.
For each of these values, 2 components characterize the association between $X$ and $Y$, initially with a positive directionality that weakens initially over time.
Then the directionality inverts and grows stronger in the other direction through components 5-8.

\begin{table}[]
  \begin{center}
    \caption{Synthetic Model Showing changing directionality over time}
    \label{tab:syn2}
    \begin{tabular}{c||c|c|c|c}
      z & $\alpha_z$ & $\lambda_z$ & $\mu_z$ & $\nu_z$ \\
      \hline
      1 & 0.125 & 1.00 & 0.0 & 0.0 \\
      2 & 0.125 & 1.00 & 1.0 & 1.0 \\
      3 & 0.125 & 0.67 & 0.0 & 0.0 \\
      4 & 0.125 & 0.67 & 1.0 & 0.5 \\
      5 & 0.125 & 0.5 & 0.0 & 0.5 \\
      6 & 0.125 & 0.5 & 1.0 & 0.0 \\
      7 & 0.125 & 0.33 & 0.0 & 1.0 \\
      8 & 0.125 & 0.33 & 1.0 & 0.0 \\
    \end{tabular}
  \end{center}
\end{table}

\begin{figure}[]
    \centering
    \begin{subfigure}[]{0.49\textwidth}
        \centering
        \includegraphics[width=\textwidth]{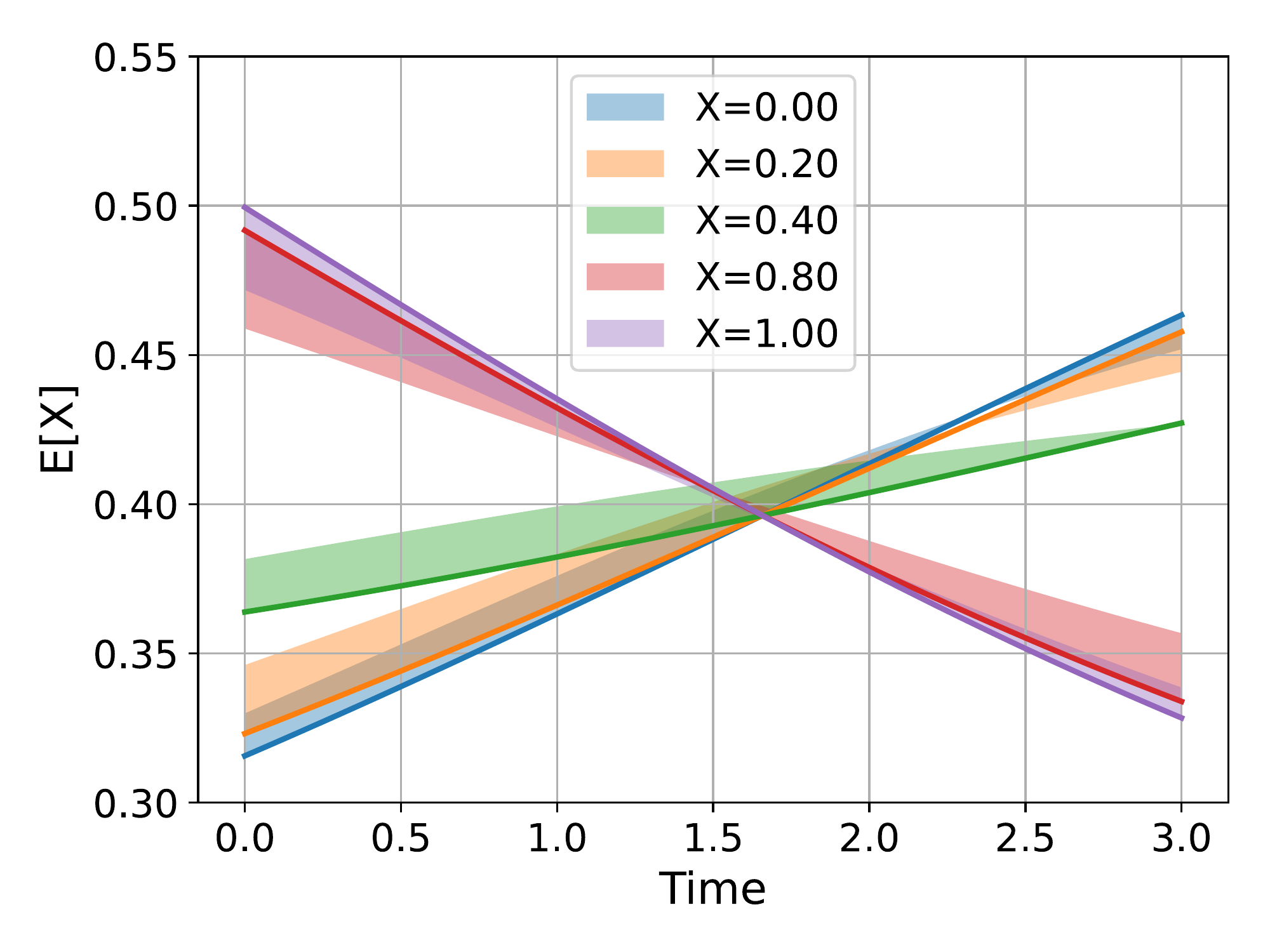}
        \caption{}
        \label{fig:synth2_time}
    \end{subfigure}
    \hfill
    \begin{subfigure}[]{0.49\textwidth}
        \centering
        \includegraphics[width=\textwidth]{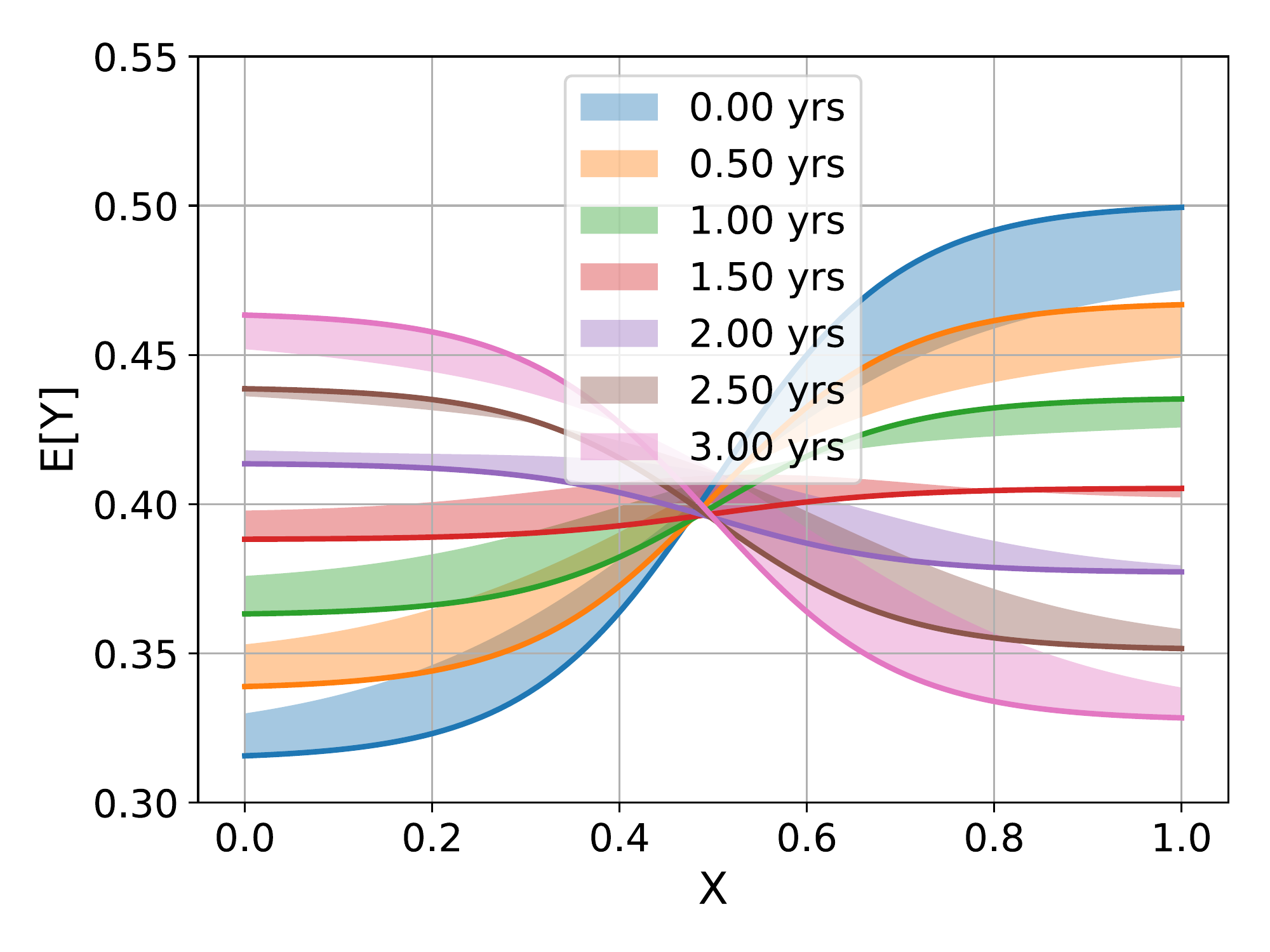}
        \caption{}
        \label{fig:synth2_x}
    \end{subfigure}
    \caption{Inference results for synthetic example with inverting association over time (Section \ref{sec:synth2}). Figure \ref{fig:synth2_time} shows the expected value of $Y$ over time for selected values of $X$. Figure \ref{fig:synth2_x} shows the same inference, but as a function of continuously valued $X$ with selected time points. For both plots, solid lines are from the original model, and shaded regions are deviations to the re-estimated model.}
    \label{fig:synth2}
\end{figure}

As in the previous example, we compute the expected value of $Y$ given $X$ over time.
Figure \ref{fig:synth2} shows these computations, both for the original model and re-estimated model.
The re-estimated model was trained on 10,000 samples drawn from the original model.

%

\subsection{Diabetes} \label{sec:res_diab}
In these results, we model the well-understood relationship between hemoglobin A1C and diabetes.
Diabetes diagnoses are typically given for \aoc{} levels greater than 6.5\%.
In this study, each $x$ is the \aoc{} value and each $\delta$ is the time elapsed to the next occurring \diab{} diagnosis ($y$ is not used).
We extracted 5,661 samples from 2,539 subjects (see Table \ref{tab:nums}).

\begin{figure}[]
    \centering
    \includegraphics[width=0.5\textwidth]{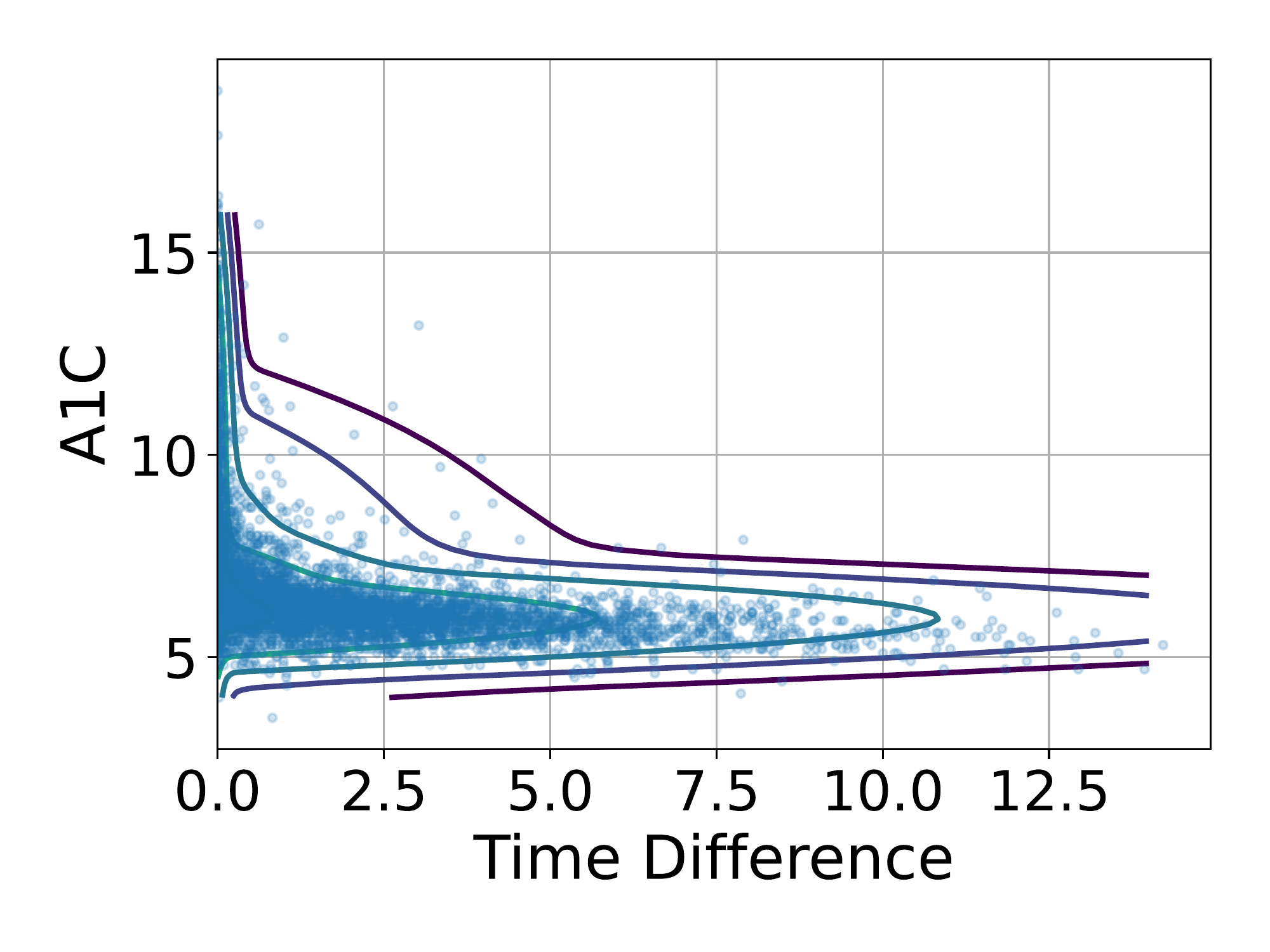}
    \caption{Scatter plot of data point used in training the \aoc{}-\diab{} model (Section \ref{sec:res_diab}) along with equiprobable contours of the trained likelihood function.}
    \label{fig:diab_cont}
\end{figure}

Running the Bayesian Optimization model selection resulted in $N_Z = 10$ components.
Figure \ref{fig:diab_cont} shows a scatter plot of the data (\aoc{} vs time interval) and equiprobable contours generated by the likelihood function of the model.
As expected, high values of \aoc{} result in a quick \diab{} diagnoses, while smaller values have a longer time interval to diagnoses.
The model captures this L-shaped distribution.

\subsection{Depression} \label{sec:res_dep}
In this section, we described results from three models trained to capture associations that \chol{}, \crp{}, and \phqt{} have with the \phqn{}.
Using the trained models, likelihood ratios are computed for moderate-severe (\phqn{} $>$ 14) vs minimal depression (\phqn{} $<$ 5) for each variable.
This is done in the two-dimensional space of measurement vs time.
These models are then combined using (\ref{eq:comp}).
Combinations of measurements and time intervals are shown that contribute to varying levels of the likelihood ratios.

\paragraph{Data extraction}
Table \ref{tab:nums} shows the number of subjects and samples extracted for each model.
One sample is a triplet that consists of a measured input ($x$), the \phqn{} score ($y$) and the time interval in years between them ($\delta$).
The measured value $x$ can be \chol{}, \crp{}, or \phqt{}, depending on the model.
Any two instances of $x$ and $y$ are included that occur in the subject's clinical record, as long as the \phqn{} score occurs after $x$ was recorded.
Extracting data in this manner means that the same value $x$ could be used in multiple samples, in cases where multiple \phqn{} scores were recorded after the $x$ value was measured.

\paragraph{Model training and comparison to baseline models}
To evaluate and compare the model fit to the baseline models, we train on half of the samples and compute the likelihood on the held-out data.
Table \ref{tab:basecomp} shows the likelihood ratios between our model (\ref{eq:modeldef}) and the two baseline models (\ref{eq:baseline1}) and (\ref{eq:baseline2}).
These values indicate how many times more likely the held-out data is for our model than the baseline model.
Thus, values greater than 1 indicate worse fit than our model.
For example, the held-out data is approximately 3 times more likely for our model than the multivariate model for \chol{} - \phqn{}.

\begin{table}[]
  \begin{center}
    \caption{Model Fit Comparison to Baseline Models}
    \label{tab:basecomp}
    \begin{tabular}{l|c|c|c}
      Model & \chol{} - \phqn{} & \crp{} - \phqn{} & \phqt{} - \phqn{} \\
      \hline
      Multivariate (\ref{eq:baseline1})         & 3.01  & 11.72  & 9.38   \\
      Conditional Bivariate (\ref{eq:baseline2})& 1.22  & 12.29  & 7.22  \\
    \end{tabular}
  \end{center}
\end{table}

\begin{figure}[]
    \centering
    \begin{subfigure}[]{0.32\textwidth}
        \centering
        \includegraphics[width=\textwidth]{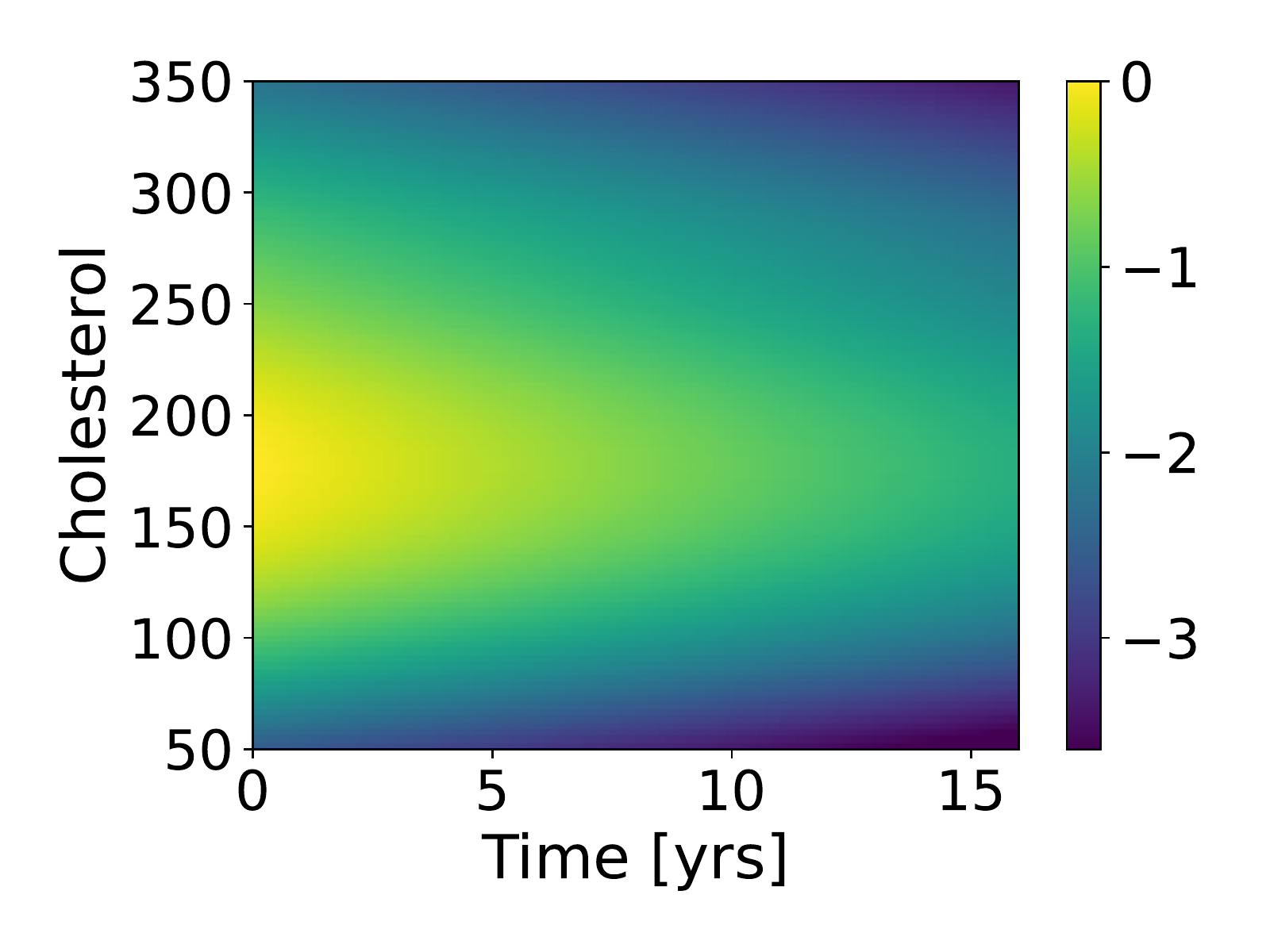}
        \caption{}
        \label{fig:chol_prev}
    \end{subfigure}
    \hfill
    \begin{subfigure}[]{0.32\textwidth}
        \centering
        \includegraphics[width=\textwidth]{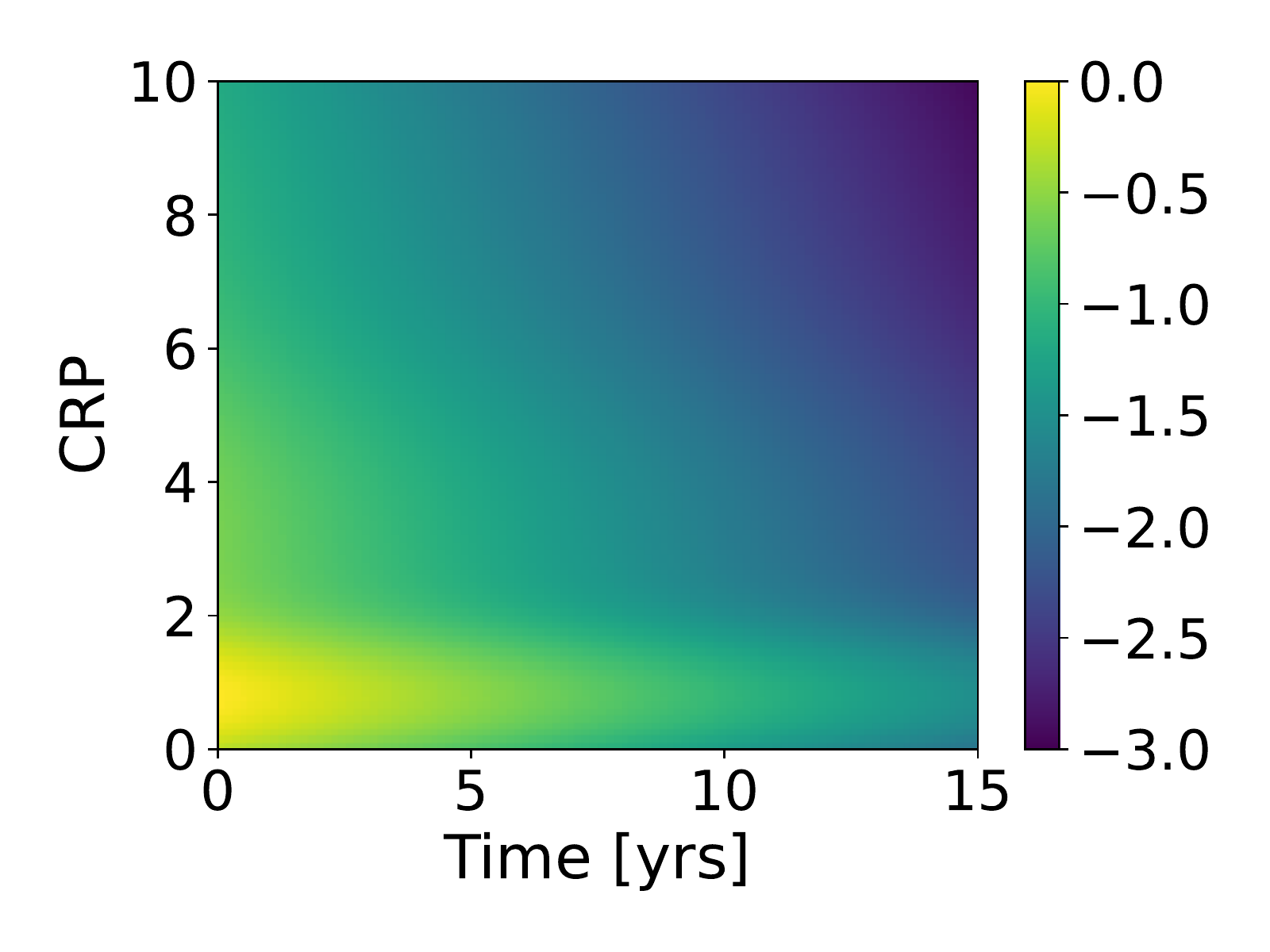}
        \caption{}
        \label{fig:crp_prev}
    \end{subfigure}
    \hfill
    \begin{subfigure}[]{0.32\textwidth}
        \centering
        \includegraphics[width=\textwidth]{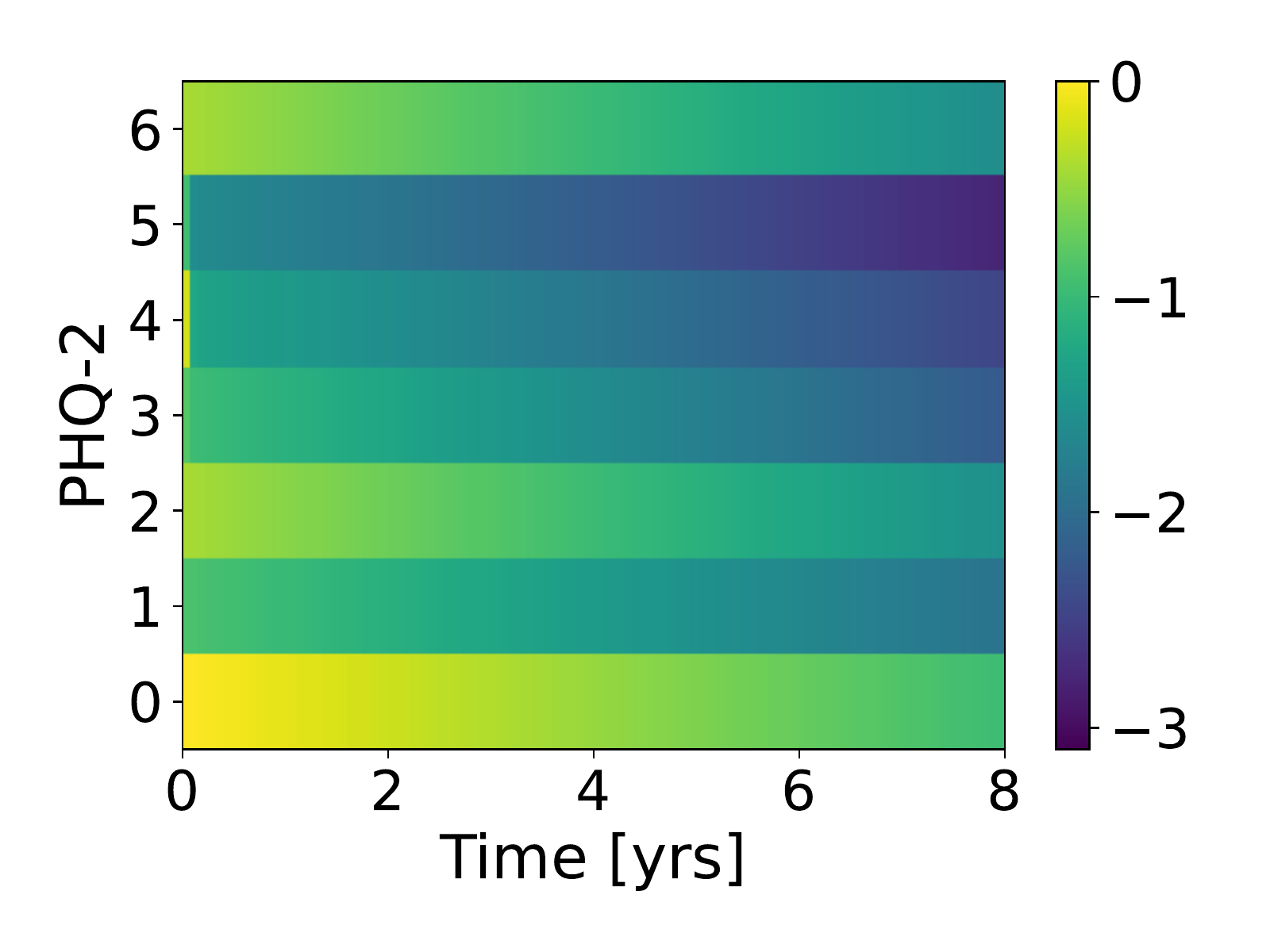}
        \caption{}
        \label{fig:phq2_prev}
    \end{subfigure}
    \hfill\begin{subfigure}[]{0.32\textwidth}
        \centering
        \includegraphics[width=\textwidth]{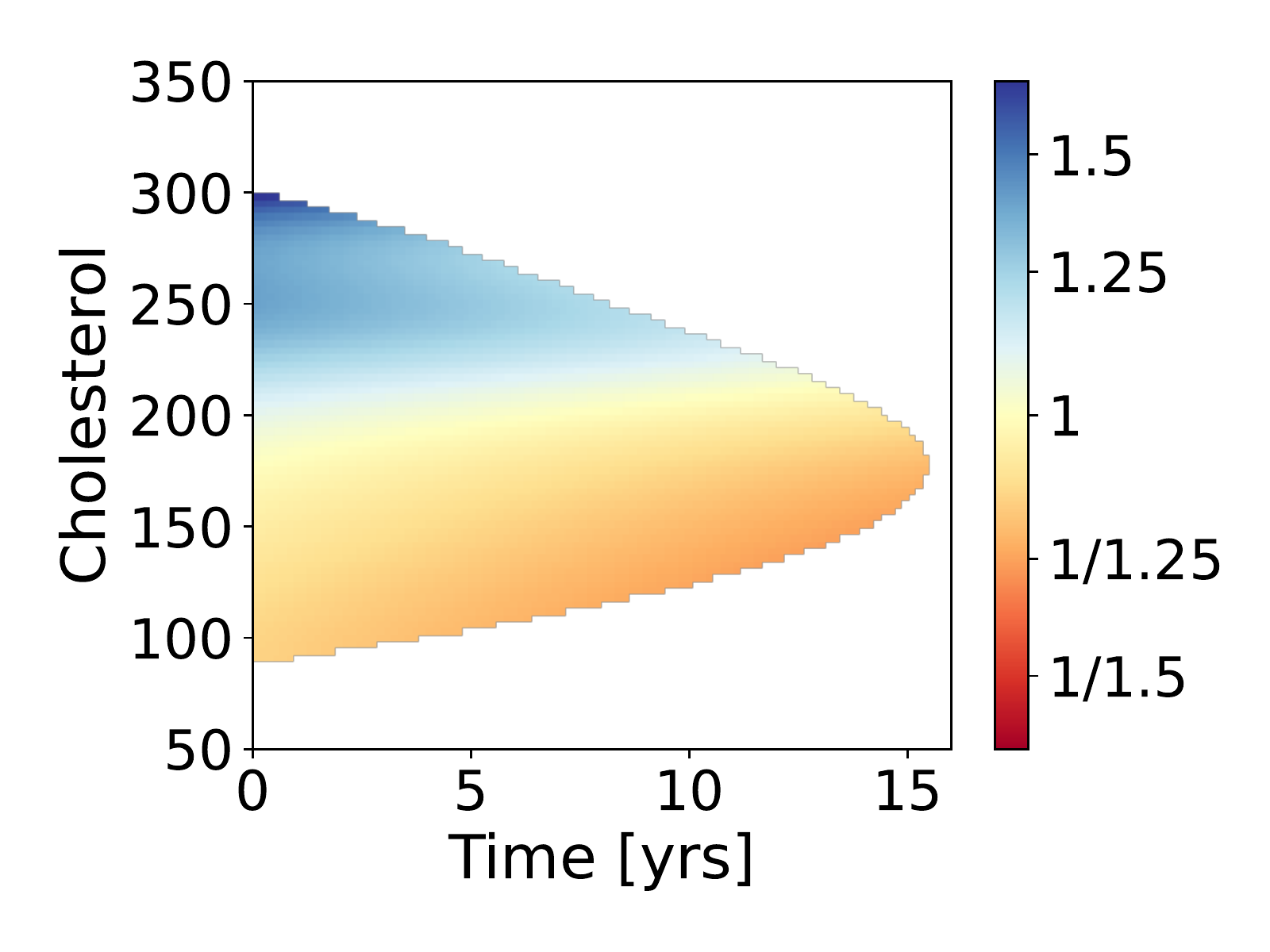}
        \caption{}
        \label{fig:chol_lr}
    \end{subfigure}
    \hfill
    \begin{subfigure}[]{0.32\textwidth}
        \centering
        \includegraphics[width=\textwidth]{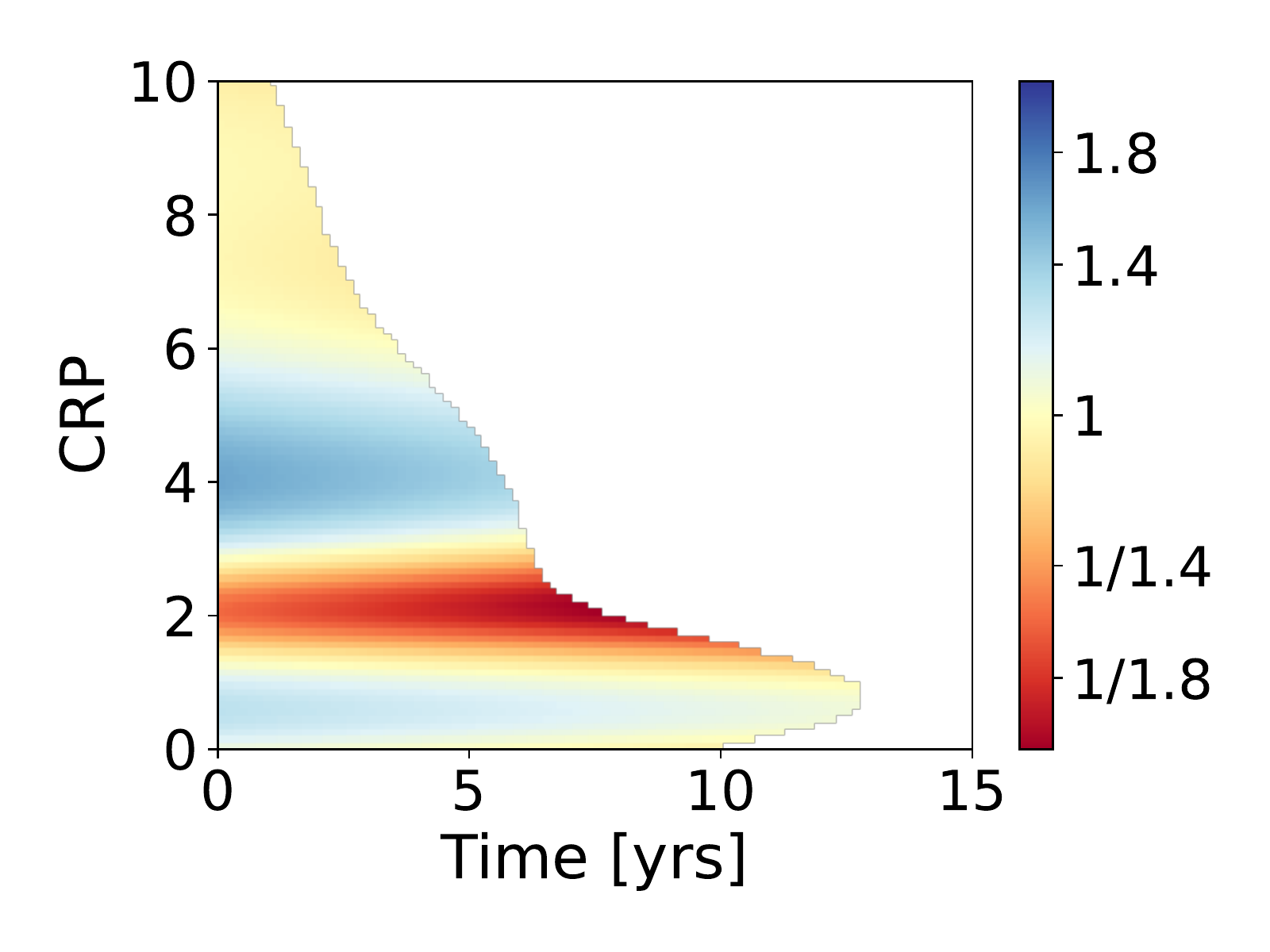}
        \caption{}
        \label{fig:crp_lr}
    \end{subfigure}
    \hfill
    \begin{subfigure}[]{0.32\textwidth}
        \centering
        \includegraphics[width=\textwidth]{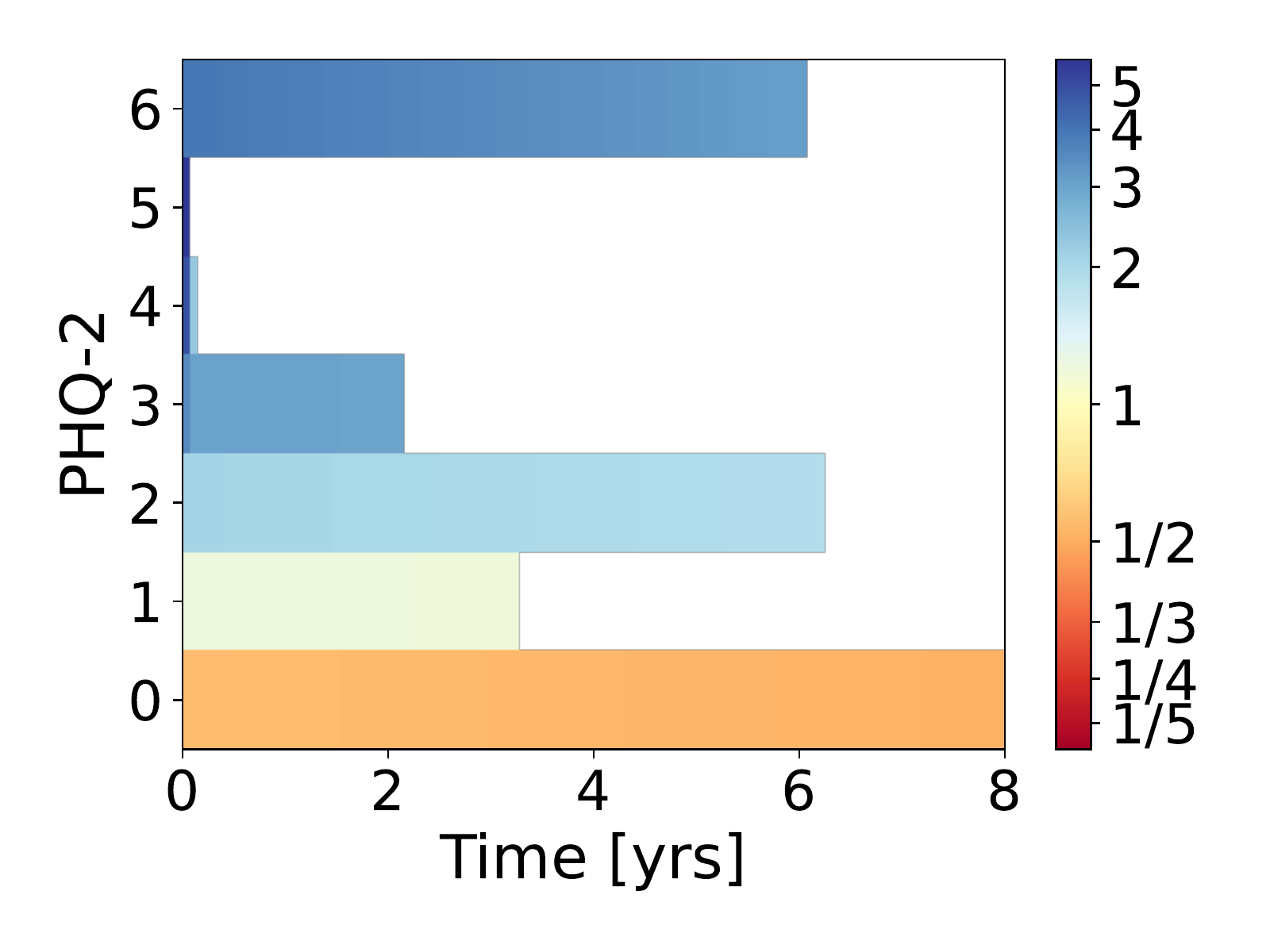}
        \caption{}
        \label{fig:phq2_lr}
    \end{subfigure}
    \caption{Normalized prevalence scores for the \chol{} (\ref{fig:chol_prev}), \crp{} (\ref{fig:crp_prev}), and \phqt{} (\ref{fig:phq2_prev}) models using a $\log_{10}$ scale. These scores are the normalized likelihoods for both the moderate-severe and minimal depression outcomes, relative to the mode. Likelihood ratios for moderate-severe vs minimal depression for these models (\ref{fig:chol_lr}, \ref{fig:crp_lr}, \ref{fig:phq2_prev}). Yellow color indicates approximately equal likelihood between outcome classes, blue indicates higher risk for the moderate-severe class, and red corresponds to a lower risk. A version for \phqt{} showing a narrow time range in shown in Figure \ref{fig:phq2_zoom}.}
    \label{fig:prev}
\end{figure}

\paragraph{Moderate-severe and minimal depression outcome classes}
Likelihood functions for two outcome classes based on the \phqn{} are defined as \phqn{} $<$ 5, 
    $$
        f_{min}\left(x, \delta\right) = \sum_{y<5}f(x, y, \delta)
    $$
for minimal depression, and \phqn{} $>$ 15, 
    $$
        f_{modsev}\left(x, \delta\right) = \sum_{y>15} f(x, y, \delta)
    $$
for moderate-severe depression.
For each model, we show (Figures \ref{fig:chol_prev}, \ref{fig:crp_prev}, \ref{fig:phq2_prev}) the total likelihood, or prevalence score, relative to its mode,
    \begin{equation} \label{eq:prev}
        p\left(x, \delta\right)=\frac{f_{min}\left(x, \delta\right) + f_{modsev}\left(x, \delta\right)}{\max \left[ f_{min}\left(x, \delta\right) + f_{modsev}\left(x, \delta\right)\right]},
   \end{equation}
which gives the likelihood of any $x,\delta$ combination relative to the most likely $x, \delta$ combination.

\begin{figure}[]
    \centering
    \begin{subfigure}[]{0.49\textwidth}
        \centering
        \includegraphics[width=\textwidth]{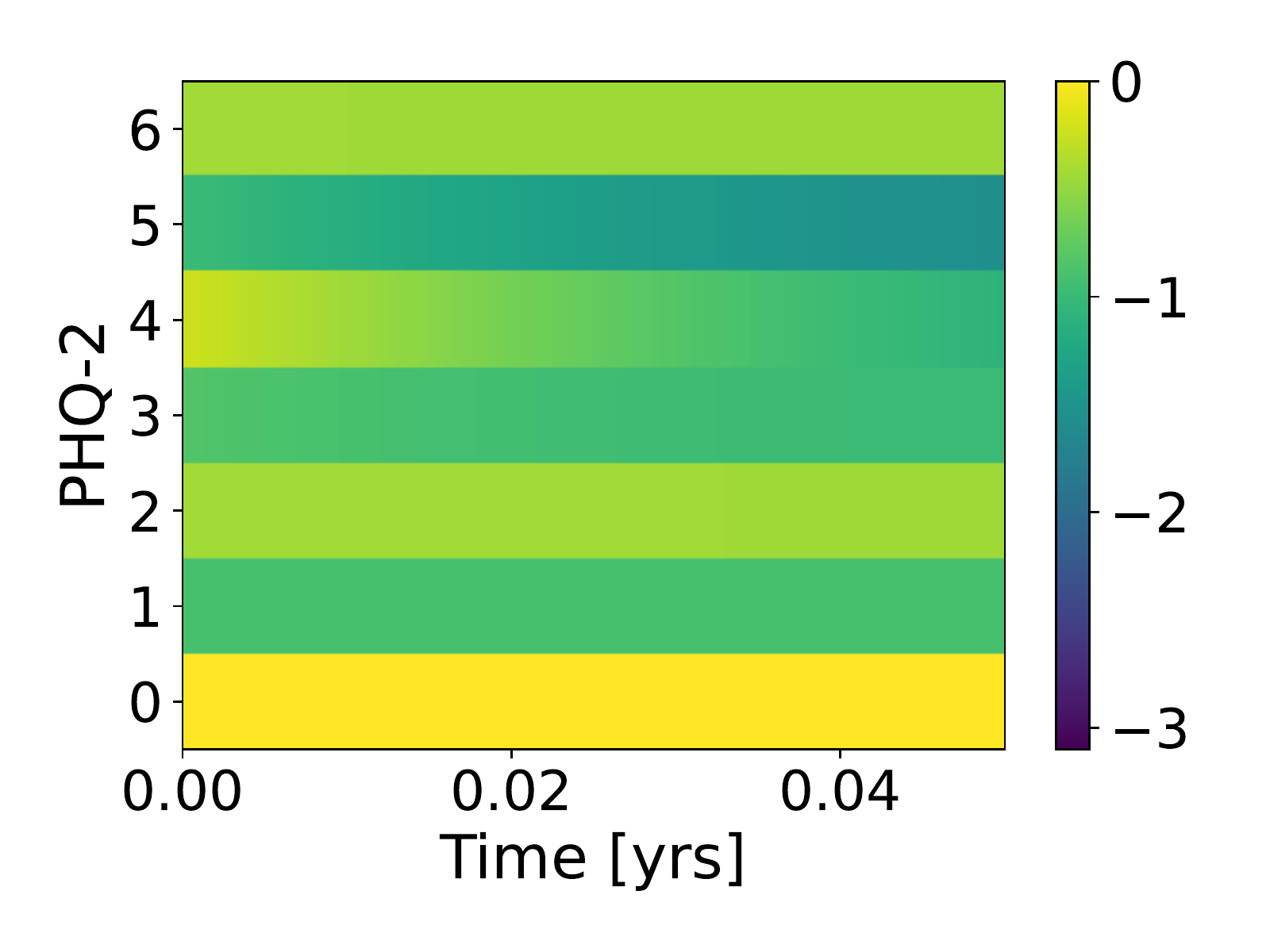}
        \caption{}
        \label{fig:phq2_zoom_prev}
    \end{subfigure}
    \hfill
    \begin{subfigure}[]{0.49\textwidth}
        \centering
        \includegraphics[width=\textwidth]{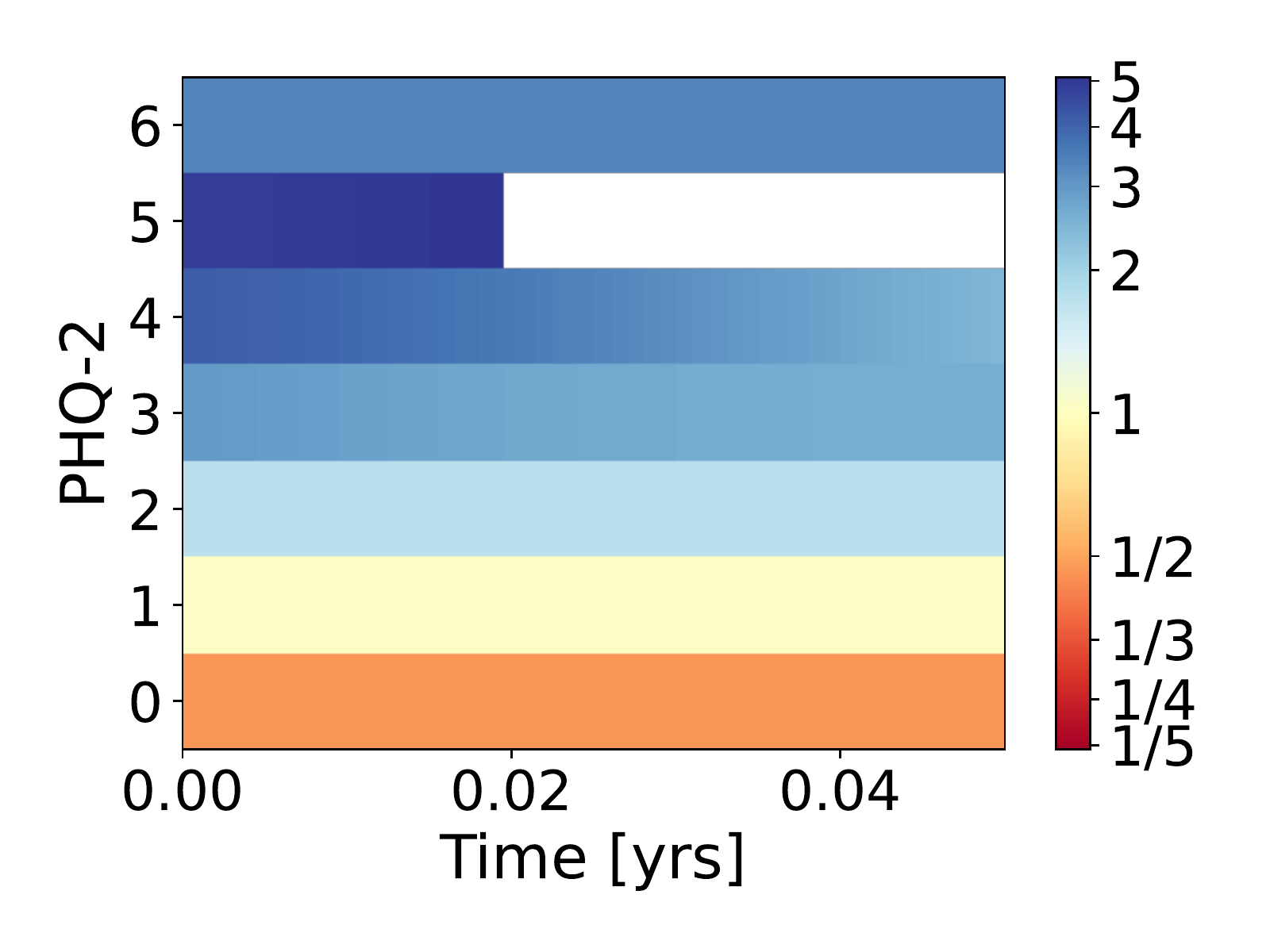}
        \caption{}
        \label{fig:phq2_zoom_lr}
    \end{subfigure}
    \caption{Prevalence scores and likelihood ratio for the \phqt{} model. Zoomed version of Figures \ref{fig:phq2_prev} and \ref{fig:phq2_lr} showing earlier times.}
    \label{fig:phq2_zoom}
\end{figure}

\paragraph{Likelihood ratio between outcome classes}
The likelihood, or odds, ratio between the two severity classes is $f_{modsev}\left(x, \delta\right)/f_{min}\left(x, \delta\right)$.
This ratio shows how much more likely moderate-severe depression is than minimal depression, as measured by the \phqn{}.
For example, a value of 2 means that the moderate-severe depression class is twice as likely as the minimal depression class, and a value of 0.5 indicates the opposite.

The model is capable of generating likelihood ratios for any continuous values of $x, \delta$, even those far outside what was seen in the training data.
In general, we should only expect to make confident assessments of likelihood for values that have a reasonable prevalence.
We use the prevalence score to help us determine appropriate regions to examine, and show (Figures \ref{fig:chol_lr}, \ref{fig:crp_lr}, \ref{fig:phq2_lr}) the likelihood ratios for the region of $x, \delta$ that has a prevalence score $p(x, \delta)$ greater than 0.05.
Figure \ref{fig:phq2_zoom} shows the prevalence score and likelihood ratios for \phqt{} in a narrow time range.

\begin{table}[]
  \begin{center}
    \caption{Number of subjects, samples, selected model order, and BIC variability due to initialization}
    \label{tab:nums}
    \begin{tabular}{l|c|c|c|c}
      Model & Subjects & Samples & Components & BIC \\
      \hline
      \aoc{}-\diab{}  & 2,539  & 5,661   & 10  & -   \\
      \chol{}-\phqn{} & 9,177  & 402,841 & 292 & 19.93 $\pm$ 0.15 \\
      \crp{}-\phqn{}  & 5,932  & 53,542  & 44  & 16.10 $\pm$ 0.05 \\
      \phqt{}-\phqn{} & 8,112  & 125,234 & 121 & 10.48 $\pm$ 0.19 \\
    \end{tabular}
  \end{center}
\end{table}

\paragraph{Composite likelihood}
These three models can be combined to evaluate the likelihood of \phqn{} over time with respect to \chol{}, \crp{}, and \phqt{}.
The likelihood function is composed using contributions from each model as described in (\ref{eq:comp}).
This results in a high-dimensional likelihood function that can be difficult to visualize.
Table \ref{tab:comp} shows sample values of time intervals and measured values that contribute towards a given composite likelihood ratio.
Each row corresponds to a likelihood ratio, and a single combination of variables that produces that odds ratio.
The first row corresponds to the smallest likelihood ratio that the models output (0.21), and the last row has the largest likelihood ratio (15.2).
Note that there are many different combinations of variables that could produce any single likelihood ratio.

\begin{table}[]
  \begin{center}
    \caption{Sample values with fixed composite likelihood ratio of moderate-severe depression vs minimal depression.}
    \label{tab:comp}
    \begin{tabular}{c|c|c|c|c|c|c}
       likelihood ratio & $\delta_\chol{}$ & $x_\chol{}$ & $\delta_\crp{}$ & $x_\crp{}$ & $\delta_\phqt{}$ & $x_\phqt{}$ \\
      \hline
      0.21 (min)    & 13.1 & 141 & 7.6  & 2.0 & 8.0 & 0 \\
      0.25          & 12.4 & 171 & 5.8  & 1.9 & 0.0 & 0 \\
      0.33          & 8.1  & 208 & 4.5  & 1.8 & 0.0 & 0 \\
      0.50          & 1.5  & 141 & 6.8  & 1.1 & 0.0 & 0 \\
      1.0           & 7.8  & 232 & 10.8 & 1.4 & 3.2 & 1 \\
      2.0           & 8.6  & 220 & 2.1  & 3.9 & 0.0 & 1 \\
      4.0           & 0.0  & 286 & 3.9  & 4.1 & 6.2 & 2 \\
      6.0           & 0.0  & 299 & 0.6  & 4.0 & 0.1 & 4 \\
      8.0           & 0.0  & 299 & 0.0  & 4.0 & 2.1 & 3 \\
      10.0          & 0.2  & 296 & 0.0  & 4.0 & 0.1 & 6 \\
      15.2 (max)    & 0.0  & 299 & 0.0  & 4.0 & 0.0 & 5 \\
    \end{tabular}
  \end{center}
\end{table}

\section{Discussion} \label{sec:disc}
The synthetic experiments show two examples of the dynamics that can be expressed with the model.
Although we chose parameters in the model in a specific way, there are many different parameter values that can lead to similar results.
In general, we do not have identifiability properties for the model.

We used 10,000 samples in these examples to show that the re-estimated models have similar behavior to the original model.
It would be possible to perform a more rigorous analysis of the estimate accuracy; however, there is also a strong dependency on the model structure.

The diabetes model was trained on extracted \aoc{} measurements and the time until the next future \diab{} diagnosis.
In this setup, several \aoc{} measurements can be extracted with respect to the same \diab{} diagnosis.
However, only the next \diab{} diagnoses is considered, not all future diagnoses.

In this problem, all of the subjects used to extract samples have \diab{} diagnoses.
This model can be used to infer how long it will take to receive a \aoc{} diagnosis, rather than if a subject will receive the diagnosis at all.
Figure \ref{fig:diab_cont} shows the raw data along with equiprobable likelihood contours of the model.
Large values of \aoc{} result in a quick \diab{} diagnosis, and smaller values result in a longer time interval to the first \diab{} diagnosis.
By visual inspection, we confirm that the model is capturing the density of samples in the data.

For the depression examples, three models were trained that include the 0-27 point \phqn{} tool.
In contrast with from the diabetes model, we extracted all pairwise instances where the \phqn{} occurred after the measured \chol{}, \crp{}, or \phqt{}.
All subjects had at least one \phqn{} instance creating a bias in the sample, as this implies that there is likely some indication of depression, or reason to test for it.

Figures \ref{fig:chol_prev} and \ref{fig:chol_lr} show the prevalence scores and likelihood ratios as a function of \chol{} and the time interval.
The normative \chol{} range is less than 200 mg/dL.
These values of \chol{} show a likelihood ratio of approximately 1 or less than 1, indicating less risk of moderate-severe depression for \chol{} less than 200 mg/dL.
On the other hand, values greater than 200 mg/dL start to show increased likelihood of moderate-severe depression relative to minimal depression.
This trend continues to approximately 300 mg/dL, given a likelihood ratio around 2, indicating that moderate-severe depression is approximately 2 times more likely.

Prevalence and likelihood ratios for \crp{} are shown in Figures \ref{fig:crp_prev} and \ref{fig:crp_lr}.
Based on the prevalence figures, a large majority of \crp{} values are in the range 0 mg/dL - 2 mg/dL.
Values greater than 6 mg/dL have likelihood ratios around 1.
The lowest likelihood ratios occur in a a range around 2 mg/dL.
The range 3 mg/dL - 6 mg/dL has the highest values.
These patterns indicate that the \crp{}-\phqn{} relationship may be more complicated than that of \chol{}-\phqn{}.

As Figures \ref{fig:phq2_prev} and \ref{fig:phq2_zoom_prev} show, prevalence for \phqt{} is mostly in very early times periods for \phqt{} $>$ 0, and more elongated through time for \phqt{} = 0.
This is reflective of the usage of the \phqt{} as a screening tool for administration of the \phqn{}.
Likelihood ratios for moderate-severe and minimal depression are shown in Figure \ref{fig:phq2_lr}.
As the \phqt{} increases, so does the risk of moderate-severe depression.
A value of \phqt{} = 0 has a likelihood ratio less than 1, whereas \phqt{} = 6 has a likelihood ratio of approximately 4.
Compared to \chol{} and \crp{}, the \phqt{} has a stronger influence on the \phqn{}.
Zooming in to short time intervals (Figure \ref{fig:phq2_zoom_lr}), we see the likelihood ratios for the less prevalent 1, 3, and 5 \phqt{} scores.

The number of components determined by the model selection procedure for each model is shown in Table \ref{tab:nums}.
This number varies for each model as the complexity required to model the data is different in each case.
The variability is due in part to the sample size, and in part due to the underlying nonlinear relationships that the model is attempting to capture.

Using all three models, we can formulate a single joint probability distribution using (\ref{eq:comp}).
This can be used to evaluate prevalence and risk for any combination of \chol{}, \crp{}, and \phqn{}.
In general, a single risk score (likelihood ratio) can be produced by many different combinations of these variables.
In Table \ref{tab:comp}, a single combination for given likelihood ratios are given.
In general, the \chol{} values increase as risk increases, but in some cases decreased \chol{} is traded for increased values in other dimensions.
For example, moving from a risk of 1.0 to 2.0, the sample provided in the Table shows \chol{} decreasing, but the \phqt{} increasing from 1 to 2.
As we have seen, the \phqt{} is a stronger indicator of \phqn{} risk than \chol{}.
A \phqt{} of 2 carries a risk of approximately 2.0 (Figure \ref{fig:phq2_lr}), requiring the other variables to have a risk around 1.0 for the total risk to be 2.0.

A situation of interest is when there are a large number of variables.
In this case, we expect many of the variables may have limited effect on the target variable $Y$.
This can be expressed through independence of a variable $X_i$ with $Y$, $f(x_i, y) = f(x_i)f(y)$.
When this occurs for a subset of the variables $X_{P+1}, X_{P+2}, \ldots X_M$, then from the model definition (\ref{eq:comp}), $f(x|x_1, \ldots x_M, \delta_1,\ldots, \delta_M) \propto  \prod_{i=1}^P f(y|x_i, \delta_i)$.
Therefore, only the dependent variables will have an impact under this model.
This analysis, however, does not include estimation error, which would contribute to inference error.

\paragraph{Limitations}
The structure defined above is designed to capture the joint distribution between variables and the time difference between them.
Within each mixture component, the choice of Gaussian distributions for the variables and an Exponential distribution for the time difference may impart model bias.
The use of mixtures of these distributions mitigates this limitation to some extent.

Estimation using the EM algorithm is sensitive to the parameter initialization.
Several approaches can be used to reduce the variability caused by this that use multiple initializations.
These include selecting the best fit model from the set of trained models or averaging resulting models.
Model selection performed by the BO method trains multiple models, each of which can be initialized differently.
Increasing a larger number of BO iterations may also help in increasing robustness.

\section{Conclusions} \label{sec:conc}
In this work, we have developed a probabilistic model that captures the joint distribution between two measurements and the time interval between them.
The model utilizes a latent variable to control the dependence between the measurements and time interval.
Estimation algorithms using EM and inference equations are derived.
A trained model can be used to calculate future risk at arbitrary time points.
In addition, multiple models can be composed to weigh the contributions of risk from each model.

We show results on data derived from the VHA.
Our focus for this work is on diabetes and depression, in particular the impact of \aoc{} on \diab{} diagnoses and that of \chol{}, \crp{}, and \phqt{} on the \phqn{}.
We show likelihood ratios for two classes of outcomes corresponding to moderate-severe and minimal depression as measured by the \phqn{} tool.

This method allows for continuous-time modeling of longitudinal EHR data.
It provides a model-based approach to calculating risk as a function of time.
In contrast to existing methods, this model is a full likelihood probabilistic model that can be used to infer any outcome class with a single model.
The model complexity can change to fit the complexity of the underlying data, as opposed to methods with a fixed number of coefficients.

There are many avenues for expanding this work.
In these results, we do not utilize any control groups.
All subjects under consideration have taken the \phqn{} in the depression models, or had a \diab{} diagnosis in the diabetes model.
It would be beneficial to train additional models using right-censored data that have never taken the \phqn{}, or have never had a \diab{} diagnosis.
This would enable evaluation of risk for not having these outcomes and thereby form a more complete picture of possible outcomes.

The composition of models does not take into account interaction between variables (for example \chol{} and \crp{}), but rather uses conditional independence of these variables given the outcome \phqn{}.
In this way, they each make conditionally independent contributions to the risk.
It is possible to expand the model into higher dimensions and taking into account higher order interactions.
However, this would greatly increase the complexity of the model and computational cost of estimating the model.
Depending on the application this may not or not be worth the added expressiveness gained.

These models can be trained on much larger collections of variables.
Since they are trained in a pairwise fashion, this can be done in a scalable manner.
The estimation of each model can also be parallelized if needed, as the EM algorithm fits neatly in the map-reduce framework.
This method is geared towards the incorporation of entire longitudinal EHR data streams that could enable probabilistic inference and quantification of risk odds for diverse subjects.
The analysis of such models is left as future work.

\section{Acknowledgements}
This work was performed under the auspices of the U.S. Department of Energy by Lawrence Livermore National Laboratory under Contract DE-AC52-07NA27344.
This research is based on data from the Million Veteran Program, Office of Research and Development (ORD), Veterans Health Administration (VHA), and was supported by award \#I01CX001729 from the Clinical Science Research and Development (CSR\&D) Service of VHA ORD.
This publication does not represent the views of the Department of Veteran Affairs or the United States Government. J.C. Beckham was also supported by a Senior Research Career Scientist Award (\#lK6BX003777) from CSR\&D.

The MVP Suicide Exemplar Workgroup for this publication includes Khushbu Agarwal, Allison E. Ashley-Koch, Mihaela Aslan, Jean C. Beckham, Edmond Begoli, Tanmoy Bhattacharya, Ben Brown, Patrick S. Calhoun, Mikaela Cashman McDevitt, Kei-Hoi Cheung, Sutanay Choudhury, Ashley M. Cliff, Judith D. Cohn, Silvia Crivelli, Leticia Cuellar-Hengartner, Haedi E. Deangelis, Michelle F. Dennis, Sayera Dhaubhadel, Patrick D. Finley, Kumkum Ganguly, Michael R. Garvin, Joel E. Gelernter, Lauren P. Hair, Phillip D. Harvey, Elizabeth R. Hauser, Michael A. Hauser, Nick W. Hengartner, Daniel A. Jacobson, Piet C. Jones, David Kainer, Alan D. Kaplan, Ira R. Katz, Rachel L. Kember, Nathan A. Kimbrel, Angela C. Kirby, John C. Ko, Beauty Kolade, John Lagergren, Matthew Lane, Daniel F. Levey, Drew Levin, Jennifer H. Lindquist, Xianlian Liu, Ravi K. Madduri, Carrie Manore, Susana B. Martins, John F. McCarthy, Benjamin H. McMahon, J. Izaak Miller, Destinee Morrow, David W. Oslin, Mirko Pavicic, John P. Pestian, Saiju Pyarajan, Xue J. Qin, Nallakkandi Rajeevan, Christine M. Ramsey, Ruy Ribeiro, Jonathon Romero, Alex Rodriguez, Daniel Santel, Noah Schaefferkoetter, Yunling Shi, Murray B. Stein, Kyle A. Sullivan, Ning Sun, Suzanne R. Tamang, Alice Townsend, Jodie A. Trafton, Angelica Walker, Xiange Wang, Victoria Wangia-Anderson, Renji Yang, Shinjae Yoo, Hong-Jun Yoon, Rafael Zamora-Resendiz, and Hongyu Zhao.

\appendix
\section{Notation} \label{app:not}
Table \ref{tab:not} contains descriptions of the notation used in this work.

\begin{table}[]
  \begin{center}
    \caption{Notation}
    \label{tab:not}
    \begin{tabular}{|c c|}
        \hline
      symbol & description \\
      \hline
        $X, Y$    & measurement random variables \\
        $\Delta$       & time elapsed between measurements random variable \\
        $x, y$    & realization of $X, Y$ \\
        $\delta$       & realization of $\Delta$ \\
        $N_Z$     & number of components \\
        $z$       & component index \\
        $f$         & joint distribution between $(\Delta, X, Y)$ \\
        $f_{exp}$   & exponential distribution \\
        $f_G$       & Gaussian distribution \\
        $\alpha_z$  & mixing coefficient for component $z$ \\
        $\boldsymbol{\alpha}$   & collection of $\alpha_1, \ldots, \alpha_{N_Z}$ \\
        $\lambda_z$ & rate parameter for component $z$ \\
        $\boldsymbol{\lambda}$  & collection of $\lambda_1, \ldots, \lambda_{N_Z}$ \\
        $\mu_z, \sigma_z^2$ & mean and variance parameters of $X$ for component $z$ \\
        $\boldsymbol{\mu}, \boldsymbol{\sigma}^2$ & collection of $\mu_i, \ldots, \mu_{N_Z}$ and $\sigma_1^2, \ldots \sigma_{N_Z}^2$ \\
        $\nu_z, \xi_z^2$    & mean and variance parameters of $Y$ for component $z$ \\
        $\boldsymbol{\nu}, \boldsymbol{\xi}^2$ & collection of $\nu_i, \ldots, \nu_{N_Z}$ and $\xi_1^2, \ldots \xi_{N_Z}^2$ \\
        $N$                 & number of samples \\
        $\delta^{(i)}, x^{(i)}, y^{(i)}$     & $i$th sample \\
        $\boldsymbol{\delta}, \boldsymbol{x}, \boldsymbol{y}$    & collection of samples $\delta^{(1)}, \ldots, \delta^{(N)}$, $x^{(1)}, \ldots, x^{(N)}$, and $y^{(1)}, \ldots, y^{(N)}$ \\
        $t$     &     time  \\
        
      \hline
    \end{tabular}
  \end{center}
\end{table}

\section{Estimation Details} \label{app:est}
The log-likelihood of the data is $L=\sum_{n=1}^N \log f(\delta_n, x_n, y_n)$.
Using the variable $k_n$ to indicate the component membership of sample $n$, the complete data log-likelihood is $\log f(\delta, x, y, k) = \sum_{n=1}^N \sum_{z=1}^Z I(k_n=z) (\log \alpha_z + \log f_{exp}(\delta; \lambda_z) + \log f_G(x; \mu_z, \sigma_z^2) + \log f_G(y; \nu_z, \xi_z^2))$.
Taking the expected value of the complete data log-likelihood with respect to $k|\delta, x, y$, we have
    \begin{equation*}
    \begin{split}
        & E_{k|\delta,x,y} \log f(\delta, x, y, k)  = \\
         & \sum_{n=1}^N \sum_{z=1}^Z \gamma_i(z) (\log \alpha_z + \log f_{exp}(\delta; \lambda_z) + \log f_G(x; \mu_z, \sigma_z^2) + \log f_G(y; \nu_z, \xi_z^2)),
    \end{split}
    \end{equation*}
where $\gamma_i(z) = f(k_i=z|\delta_i,x_i,y_i)$.
The estimation algorithm (Algorithm \ref{alg:est}) iterates between calculating $\gamma_i(z)$ and maximizing the expected complete data log-likelihood function.

\section{Bayesian Optimization Parameters} \label{app:bo}
We use the Bayesian Optimization Python package \cite{Nogueira2014-ds}.
The bounds on $Z$ are initially set to $[0, 1000]$.
Sequential domain reduction is used to shrink the domain as the iterations progress \cite{Stander2002-gy}.
The default values of the parameters are used: shrinkage=0.7, panning=1, and zoom=0.9).
We start with 4 initial points and proceed with 20 iterations.


\bibliography{mybibfile}

\end{document}